\documentclass[prl,showpacs,a4paper,superscriptaddress,floatfix,notitlepage]{revtex4}
\usepackage{graphicx}
\usepackage{hyperref}
\usepackage{oldgerm}
\usepackage{color}
\usepackage{bbm}

\newcommand{\bra}[1]{\left\langle #1\right|}
\newcommand{\ket}[1]{\ensuremath{|#1\rangle}}

\newcommand{\bea}{\begin{eqnarray}}
\newcommand{\eea}{\end{eqnarray}}

\newcommand{\rE}{\mathrm{E}}
\newcommand{\rGME}{\mathrm{GME}}
\newcommand{\rA}{\mathrm{A}}
\newcommand{\rB}{\mathrm{B}}
\newcommand{\rf}{\mathrm{f}}
\newcommand{\rmax}{\mathrm{max}}
\newcommand{\rmin}{\mathrm{min}}
\newcommand{\ri}{\mathrm{i}}
\newcommand{\ba}{\begin{eqnarray}}
\newcommand{\ea}{\end{eqnarray}}
\newcommand{\Tr}{\text{Tr}}

\newcommand{\red}{}
\newcommand{\blk}{\color{black}}
\providecommand{\openone}{\leavevmode\hbox{\small1\kern-3.8pt\normalsize1}}

\begin{document}
\title{Thermodynamic cost of creating correlations}
\author{Marcus Huber}
\affiliation{Departament de F\'{i}sica, Universitat Aut\`{o}noma de Barcelona, E-08193 Bellaterra, Spain}
\affiliation{ICFO-Institut de Ci\`encies Fot\`oniques, Mediterranean Technology Park, 08860 Castelldefels (Barcelona), Spain}
\author{Mart\'i Perarnau-Llobet}
\affiliation{ICFO-Institut de Ci\`encies Fot\`oniques, Mediterranean Technology Park, 08860 Castelldefels (Barcelona), Spain}
\author{Karen V. Hovhannisyan}
\affiliation{ICFO-Institut de Ci\`encies Fot\`oniques, Mediterranean Technology Park, 08860 Castelldefels (Barcelona), Spain}
\author{Paul~Skrzypczyk}
\affiliation{ICFO-Institut de Ci\`encies Fot\`oniques, Mediterranean Technology Park, 08860 Castelldefels (Barcelona), Spain}
\author{Claude Kl\"ockl}
\affiliation{Departament de F\'{i}sica, Universitat Aut\`{o}noma de Barcelona, E-08193 Bellaterra, Spain}
\author{Nicolas Brunner}
\affiliation{D\'epartement de Physique Th\'eorique, Universit\'e de Gen\`eve, 1211 Gen\`eve, Switzerland}
\author{Antonio Ac\'in}
\affiliation{ICFO-Institut de Ci\`encies Fot\`oniques, Mediterranean Technology Park, 08860 Castelldefels (Barcelona), Spain}
\affiliation{ICREA-Instituci\'o Catalana de Recerca i Estudis Avan\c cats, Lluis Companys 23, 08010 Barcelona, Spain}

\begin{abstract}
We investigate the fundamental limitations imposed by thermodynamics for creating correlations. Considering a collection of initially uncorrelated thermal quantum systems, we ask how much classical and quantum correlations can be obtained via a cyclic Hamiltonian process. We derive bounds on both the mutual information and entanglement of formation, as a function of the temperature of the systems and the available energy. While for a finite number of systems there is a maximal temperature allowing for the creation of entanglement, we show that genuine multipartite entanglement---the strongest form of entanglement in multipartite systems---can be created at any temperature when sufficiently many systems are considered. This approach may find applications, e.g. in quantum information processing, for physical platforms in which thermodynamic considerations cannot be ignored. 
\end{abstract}
\maketitle

\section{Introduction}
Thermodynamics is intimately connected to information theory. In recent years, this connection has been explored and extended in the quantum world \cite{koji}. Making use of the concepts and tools of quantum information theory, this research line brought tremendous progress in our understanding of the thermodynamics of quantum systems, see e.g. \cite{lydia,spekkens,brandao,oppi}. Given the power of quantum information processing, it is natural to investigate the possibilities offered by quantum effects (such as coherence and entanglement) in the context of thermodynamics \cite{alicki,karen,nicolas,lutz,correa, rodi,lutz2,mauro,valerio,jens}.

The main question explored in this work is the following: what is the thermodynamic cost of establishing classical and quantum correlations? Our goal here is to find what are the fundamental limitations imposed by thermodynamics for creating correlations.

Here we investigate these issues using a particularly simple model. We consider a thermally isolated quantum system, composed of two (or more) uncorrelated subsystems, all initially in a thermal state. In order to establish correlations between the subsystems, we allow ourselves to perform any possible unitary operation on the entire system. Performing such a unitary will in general cost us some energy.

The first set of questions we seek to answer is how the temperature of the initial state limits the ability to create different types of correlations in the system, starting with classical correlations in bipartite and multipartite systems, before moving onto bipartite entanglement and then different forms of entanglement in the multipartite case, including the strongest form -- genuine multipartite entanglement. In all cases we provide explicit protocols for generating correlations. For arbitrarily large initial temperatures one is able to produce classical correlations, i.e. there is no threshold temperature. For entanglement, in bipartite systems we find the threshold, but show that for every type of multipartite entanglement the threshold can be made arbitrarily large by considering a sufficiently large number of subsystems. We finally exhibit upper bounds on the threshold temperature, which show that our protocols perform almost optimally, achieving the same scaling behaviour as the bound. 

After having established the bounds imposed by the temperature, we then move on to the question of how the available energy limits the correlations, by determining the maximal amount of correlation that can be created given access to a limited amount of energy. Here our focus is primarily on the bipartite setting, where we investigate optimal protocols for generating classical correlations and bipartite entanglement with limited energy. 

These results demonstrate the limitations on creating correlations that arise from thermodynamics. We envisage therefore that they will be relevant for discussing quantum information tasks in physical systems where thermodynamic considerations cannot be ignored. Similar issues were raised in NMR \cite{chuang,braunstein} and in connection with non-cyclic unitary dynamics of two particle entanglement \cite{lutz2,jens} and its work cost \cite{lutz2} in harmonic chains. From a more theoretical point of view, our results establish a link between fundamental resources of two theories:  entanglement theory \cite{horo,chrisjens} and the resource theory of thermodynamics \cite{spekkens,oppi}.

\section{Framework}
We consider a system of $n$ initially uncorrelated $d$-dimensional quantum subsystems. Each subsystem is taken to have the same (arbitrary) local Hamiltonian $H$, and the same temperature $k_\rB T= 1/\beta$. Hence the initial state of the system is 
\begin{equation}
\rho_i=\tau_\beta ^{\otimes n}, \hspace{10mm} {\rm where} \hspace{2mm} \tau_\beta =                \frac{e^{-\beta H}}{\mathcal{Z}}
\end{equation}
and $\mathcal{Z} = \Tr\left(e^{-\beta H}\right)$ is the partition function. When discussing qubits we will denote by $E$ the energy of the excited state and 
\begin{equation}
p=\frac{1}{1+e^{-\beta E}}
\end{equation}
the ground state probability. Allowing ourselves the use of arbitrary (global) unitaries $U$ acting on the system, we want to characterise (i) what are the limitations imposed by the initial temperature on the available correlations (either classical or quantum) (ii) what is the energy cost $W$ of creating correlations, where $W$ is given by 
\begin{equation}
W = \Tr\left(H_\mathrm{tot}( \rho_\rf - \rho_i)\right),
\label{Wcost}
\end{equation}
where $\rho_\rf=U\tau_\beta^{\otimes n} U^\dag$ is the final state and $H_\mathrm{tot} = \sum_i H^{(i)}$ is the total Hamiltonian.  We end by noting that here, since we are interested in fundamental limitations arising from thermodynamics alone, we consider the most general operations possible, that of arbitrary global unitaries.  We will discuss this point further in the conclusions, as well as the prospects of going beyond it in future work. 

\section{Limitations arising from the temperature}
In the first half of this paper we will consider the question of how the temperature of the initial state affects the amount of correlation or entanglement that we can be created. In particular, we will impose only the minimal requirement that the processing be a unitary one, and will not ask for further constraints, either in terms of the energy cost of the process, or the efficiency of the implementation. As such, the results presented here will constitute fundamental limits on the creation of correlation or entanglement which arise solely from the thermal nature of the initial states, and their corresponding temperature.

We will first consider the creation of correlations, both in the bipartite and multipartite settings, before moving on to the question of entanglement generation, again in both the bipartite and multipartite settings.

\subsection{Correlations}

\subsubsection{Bipartite systems}

Let us start by considering the case of a two qudit system, i.e. two $d$-level systems. Correlations between the two subsystems (which shall be referred to as $A_1$ and $A_2$) can naturally be measured using the quantum mutual information $I(A_1:A_2)$
\begin{equation}
I(A_1:A_2) = S(A_1) + S(A_2) - S(A_1A_2),
\label{MI}
\end{equation}
where $S(X) = -\Tr(\rho_X \log \rho_X)$ is the von Neumann entropy of system $X$. 

The goal is then to find the the optimal unitary operation $U$ such that $\rho_\rf=U\tau_\beta \otimes \tau_\beta U^\dag$ has the maximal possible mutual information. Note first that initially $I(A_1 : A_2) = 0$, as the initial state factorizes. Thus, to create correlations, one must find a global unitary that increases the local entropies $S(A_i)$ of $\rho_\rf$, since the total entropy $S(A_1A_2) = 2S(\tau_\beta)$ cannot change. Since for a $d$-level system the local entropy is upper bounded by $S(A_i) \leq \log d$, the maximal possible mutual information is upper bounded by 
\begin{equation}\label{Imax2}
I(A_1:A_2) \leq 2 [ \log d - S(\tau_\beta)].
\label{Imax}
\end{equation}
This bound can always be achieved, by making use of the following protocol, which amounts to rotating from the energy eigenbasis to the generalized Bell basis, i.e. to a basis of maximally entangled qudit states. In more detail, for all $d$ one can define the unitary operators 
\begin{eqnarray}
X &= \sum_{m} \ket{m+1\bmod d}\bra{m},\qquad Z &= \sum_m \omega^m \ket{m}\bra{m},
\end{eqnarray}
with $\omega = e^{2\pi i/d}$ as generalisations of the (qubit) Pauli operators $\sigma_x$ and $\sigma_z$. The Bell basis $\{ \ket{\phi_{ij}}\}_{ij}$ is then given by 
\begin{equation}
\ket{\phi_{ij}} = Z^i \otimes X^j  \ket{\phi},
\end{equation}
where $\ket{\phi} = \frac{1}{\sqrt{d}}\sum_i \ket{ii}$. Finally, we consider the operation given by
\begin{equation}
U = \sum_{ij} \ket{\phi_{ij}}\bra{ij}.
\end{equation}
Since the initial state is a mixture of energy eigenstates, $\rho_\rf$ is a mixture of Bell states. Finally, since these all have maximally mixed marginals, i.e. $\Tr_{A_k} (\ket{\phi_{ij}}\bra{\phi_{ij}}) = \openone/d$ the bound (\ref{Imax2}) is achieved. We end by noting that the maximally mixed state $\openone/d$ corresponds to the infinite-temperature thermal state $\tau_0$. We shall see in the second half of the paper that when one has in addition a constraint on the energy, that the  optimal protocol produces thermal marginals, only there at lower temperatures. 

Finally, we note that for all finite initial temperatures $\beta \neq 0$ the mutual information that can be created between the two subsystems is non-zero, i.e. that one can produce correlations between them at arbitrary finite temperatures. 

\subsubsection{Multipartite systems}

In the multipartite setting one can generalise the notion of mutual information by considering the difference between the sum of local entropies and the total entropy of the system. That is, for a collection of $n$ subsystems $A_1, \ldots, A_n$, we define the multipartite mutual information as
\begin{equation}
I(\{A_i : \cdots : A_n) = \sum_{i=1}^n S(A_i) - S(A_1\cdots A_n),
\label{Igen}
\end{equation}
which vanishes only when the total system is a direct product. Again, since the total entropy of the system is conserved, to maximise this quantity one must maximise the sum of final local entropies after the protocol. The analogous upper bound,
\begin{equation}
I(\{A_i : \cdots : A_n) \leq n(\log d - S(\tau_\beta)),
\end{equation}
is seen to hold, and can again be achieved by rotating the energy eigenbasis to a basis of generalised GHZ states. Namely, one can define the basis $\{\ket{\phi_{i_1\cdots i_n}^{n}}\}_{i_1\cdots i_n}$ by
\begin{equation}
\ket{\phi_{i_1\cdots i_n}^n} = Z^{i_1} \otimes X^{i_2} \otimes \cdots \otimes X^{i_n} \ket{\phi^n},
\end{equation}
where $\ket{\phi^n} = \frac{1}{\sqrt{d}}\sum_i \ket{i}^{\otimes n}$ and the operation given by
\begin{equation}
U = \sum_{i_i,\ldots,i_n} \ket{\phi_{i_1\cdots i_n}}\bra{i_1 \cdots i_n}.
\end{equation}
Again, since the final state of the system is a mixture of generalised GHZ states, all of which have maximally mixed marginals $\Tr_{\overline{A}_k} (\ket{\phi_{ij}}\bra{\phi_{ij}}) = \openone/d$ (where $\overline{A}_k$ denotes tracing over all subsystems except $A_k$) the bound is seen to be saturated. Finally, as long as the initial temperature is not infinite $\beta \neq 0$, then the bound is non-zero, and a finite amount of correlation can be created between all of the subsystems.

\subsection{Entanglement}
Having seen in the previous section that it is possible to create correlations among the subsystems of a general multipartite system starting at arbitrary temperatures in a relatively easy fashion, we now move on to the move interesting question of creating entanglement. We will first look at the case of bipartite systems, where there is a single notion of entanglement, before moving on to multipartite systems, where there are a number of inequivalent notions of entanglement that we will study. In all cases we will restrict ourselves to the study of qubits. 

\subsubsection{Bipartite systems}
We shall start our study of the bipartite case with the simplest possible scenario, involving two qubits. Although there is only a single notion of entanglement, one can nevertheless define many inequivalent measures of entanglement. Here for concreteness we will focus on the \emph{concurrence} \cite{wootters}, which for pure states is the linear entropy of the reduced state of one party, 
\begin{equation}
C(\psi) = \sqrt{2(1-\Tr(\rho_\rA^2))},
\end{equation}
where $\rho_\rA = \Tr_\rB \ket{\psi}\bra{\psi}_{\rA\rB}$, and is extended to mixed states via the convex-roof construction
\begin{equation}
C(\rho) = \inf \sum_i p_i C(\psi_i),
\end{equation}
where the infimum is taken over all pure state decompositions $\rho = \sum_i p_i \ket{\psi_i}\bra{\psi_i}$. The concurrence is important as for qubits the convex roof can be analytically calculated and the \emph{entanglement of formation} \cite{wootters} is functionally related to it.

Crucially, for our purposes the problem of finding the state of maximal concurrence given only its spectrum was solved in \cite{ishizaka,verstraete}, which is an alternative way of phrasing the problem which we are interested in here. Moreover, it was shown that the optimal protocol not only maximises the concurrence (and therefore the entanglement of formation), but also two other important measures of entanglement, the relative entropy of entanglement, and the negativity. 

The protocol of \cite{verstraete} is easiest understood by decomposing it into a product of two unitaries, $U = V_2V_1$, where $V_1$ is a CNOT gate
\begin{equation}
V_1 = \ket{00}\bra{00} + \ket{01}\bra{01} + \ket{11}\bra{10} + \ket{01}\bra{11},
\end{equation}
and $V_2$ is a rotation in the subspace spanned by $\{\ket{00}, \ket{11}\}$ to maximally entangled states 
\begin{equation}
V_2 = \ket{\phi_{00}}\bra{00} + \ket{01}\bra{01} + \ket{10}\bra{10} + \ket{\phi_{10}}\bra{11}.
\label{V_2}
\end{equation}
Denoting by $\{\lambda_i\}_i$ the eigenvalues of the initial state $\rho_\ri$ arranged in non-increasing order, the  concurrence of the final state $\rho_\rf = V_2V_1\rho_\ri V_1^\dagger V_2^\dagger$ is given by 
\begin{equation}
C = \max (0, \lambda_1 - \lambda_3 - 2\sqrt{\lambda_2 \lambda_4}).
\label{Cmaxtwoqubits}
\end{equation}
Applied to the case at hand, with $\rho_\ri = \tau_\beta \otimes \tau_\beta$ we finally obtain
\begin{equation}
	C_{\mathrm{max}} =\max(0, 2p^2 - p - 2(1-p)\sqrt{p(1-p)}).
\end{equation}
It follows therefore, that unlike when considering correlations, there is a now a threshold temperature, $k_\rB T_\rmax /E \approx 1.19$ (or equivalently a threshold ground-state population $p_\rmin \approx 0.698$), such that for all $T \geq T_\rmax$ (or $p \leq p_\rmin$) no entanglement can be created between the two qubits.


\subsubsection{Multipartite systems} We now switch our attention to the multipartite setting. Here we will see that the limiting temperature $T_\rmax$ below which one can create entanglement can be increased when several copies of the system are jointly processed. Essentially, as more copies are available, the global system contains larger energy gaps and thus subspaces with higher purity, which can then potentially be more easily entangled. In the following we make this intuition precise by studying the dependence of $T_{ \rm max}$ on the number of copies $n$. At the same time, we study several classes of entanglement that naturally appear in the multipartite case including its strongest form: genuine multipartite entanglement. 

\emph{Entanglement in all bipartitions.} 

To start our discussion, we consider the case of $n$ qubits and a straightforward generalization of the above two-qubit protocol. That is, we consider a rotation in the $\ket{0}^{\otimes n}$, $\ket{1}^{\otimes n}$ subspace, of the form (\ref{V_2}),
\ba U=\ket{\phi^n}\bra{0}^{\otimes n} + \ket{\phi^{n'}}\bra{1}^{\otimes n} + \openone -(\ket{0} \bra{0})^{\otimes n} -(\ket{1} \bra{1})^{\otimes n}
\label{Ucreation}
\ea 
where $\ket{\phi^{n'}}=\ket{\phi^n_{10\cdots 0}}$. For a given bipartition $j|n-j$ (i.e. a partition of $j$ qubits vs. $n-j$ qubits), the concurrence in the final state $\rho_f$ can be conveniently lower bounded using the relation \cite{marcus}
\begin{eqnarray} C &\geq& 2 \Big( \left| \bra{0}^{\otimes n} \rho_f \ket{1}^{\otimes n} \right| 
\\ &&- \sqrt{\bra{0}^{\otimes j} \bra{1}^{\otimes (n-j)} \rho_f\ket{0}^{\otimes j} \ket{1}^{\otimes (n-j)}}\sqrt{\bra{1}^{\otimes j} \bra{0}^{\otimes (n-j)} \rho_f\ket{1}^{\otimes j} \ket{0}^{\otimes (n-j)} } \Big)\nonumber
\label{Cmulti}
\end{eqnarray}
and due to the simple form of $\rho_f$, these bounds are in fact tight \cite{rafsanjani}. 
Evaluating explicitly, we then obtain 
\begin{equation}
C=\lambda_0 -\lambda_n -2\sqrt{\lambda_j \lambda_{n-j}}, \hspace{10mm} \lambda_j=\bra{0}^{\otimes n-j}\bra{1}^{\otimes j} \rho_i  \ket{0}^{\otimes n-j}\ket{1}^{\otimes j}.
\label{Cbip}
\end{equation}
which is independent of the bipartition, and given by 
\begin{equation}
C =p^n-(1-p)^n-2p^{n/2}(1-p)^{n/2}.
\end{equation}
By demanding $C>0$, we can characterise the smallest $p$, and thus the largest $T$, that allows for entanglement to be created simultaneously across all bipartitions, as a function of $n$. We find a linear scaling in $n$ for this critical temperature $T_\rE^{\rm(all\hspace{1mm} bip.)}$, 
\begin{equation}
\frac{k_\rB T_{\rE}^{\rm(all\hspace{1mm} bip.)}}{E}\geq\frac{n}{2\ln(1+\sqrt{2})}.
\label{Teall}
\end{equation}
Hence it follows that entanglement across all bipartitions can always be generated starting from an arbitrary temperature $T$, by considering a sufficiently large number of qubits $n$. We note also that if one used instead of concurrence the negativity across a bipartition, a straightforward calculation shows that the same bound is obtained. 

\emph{Entanglement in a single bipartition.} 

The above protocol can be improved if the aim is to generate entanglement in a given single bipartition $j|n-j$. As in the two-qubit protocol, the idea is to perform a permutation of the initial diagonal elements before applying the rotation (\ref{Ucreation}). From expression (\ref{Cbip}), we see that the optimal permutation is the one where $\lambda_0=p^n$, $\lambda_n=\lambda_j=p(1-p)^{n-1}$ and $\lambda_{n-j}=(1-p)^n$.
In such a case, we a similar analysis to above leads to the limiting temperature, which, for large $n$ is given by  
\ba 
\frac{k_\rB T_\rE}{E}\gtrsim\frac{n-1/2}{\ln(3)}.
\label{T_E}
 \ea 
Hence the threshold temperature for the creation of bipartite entanglement using this protocol is also linear in $n$ (for high temperatures), but improves upon the above protocol in the constants. Thus for fixed $n$, one can generate entanglement across a single bipartition for slightly higher temperatures. 

\emph{Genuine multipartite entanglement}

Genuine multipartite entanglement (GME) is the strongest form of entanglement in multipartite systems. A state $\rho$ is GME iff it only admits decompositions of the form
\begin{equation}
\rho=\sum_i p_i |\phi_i \rangle \langle \phi_i |
\end{equation}
where at least one $|\phi_i \rangle $ is entangled in every possible bipartition. It follows that a necessary but not sufficient condition for GME is that $\rho$ itself is entangled across every bipartition. This suggests that the previously considered protocol for generating entanglement in all bipartitions is a natural candidate to gain a first insight on the maximal temperature for GME creation.

After applying the unitary (\ref{Ucreation})c, the state $\rho_f$  is essentially \red a GHZ-state mixed with (diagonal) noise\blk. For such a simple form, the techniques of Ref. \cite{marcus,mamaju} give us the necessary and sufficient conditions for the creation of GME \cite{rafsanjani2}, namely
\begin{equation}
\rho_{f} \hspace{1mm} {\rm is} \hspace{1mm} {\rm GME}  \iff p^n-(1-p)^n-2(2^{n-1}-1)p^{n/2}(1-p)^{n/2}\geq 0
\end{equation}
This condition leads to a lower bound on the threshold temperature for creating GME, $T_{\rm GME}$, which turns out to be asymptotically independent of $n$, and given by 
\begin{equation}
\frac{k_\rB T_{\rGME}^{\rm (GHZ)}}{E} \simeq \frac{1}{2\ln(2)}.
\end{equation}
where we added the suffix GHZ because the target entangled state of this protocol is a GHZ state. \red Moreover, as we show in the appendix, this result holds for all states whose density matrix features only diagonal and anti-diagonal elements, also known as X-states \cite{Yu,MarcusX}\blk. 

\emph{Genuine multipartite entanglement II}

Recall that there are many inequivalent types of multipartite entangled states and GHZ states only constitute one prominent class. \blk In fact it is much more favorable to use protocols that target another type of entangled states, \red namely Dicke states \cite{dicke}. An $n$-qubit Dicke state with $k$ excitations is defined as: 
\begin{equation}
|D^n_k \rangle = \frac{1}{\sqrt[]{n \choose k}]}\sum_j P_j\{|1\rangle^{\otimes k} |0\rangle^{\otimes n-k}\}
\label{dickeK}
\end{equation}
where $\sum_j P_j\{ \}$ is a sum over all possible permutations. Besides being relevant for the theory of light-matter interaction, Dicke states are useful for various quantum information tasks \cite{chiuri}, have been detected experimentally \cite{haffner,wieczorek} and have shown to exhibit  genuine multipartite entanglement \cite{duan,lucke}. 

By constructing a protocol that uses the state (\ref{dickeK}) as the target entangled state, we will show that the threshold temperature for generating GME is given by 
\ba\label{e:TGME}
\frac{k_\rB T_{\rGME}}{E} \geq\frac{n}{(k+1)\ln n}+\mathcal{O}\left[\frac{n}{(\ln n)^2}\right].
\label{Tgme}
\ea
The scaling is almost linear with $n$, which allows now for the creation of GME for an arbitrarily high temperature $T<\infty$, by considering a sufficient number of qubits $n$. Note that this result is quite counter-intuitive, as the complexity of the task we consider, entangling all qubits, increases with $n$. Furthermore, it in stark contrast with the results obtained above for the GHZ class, and thus indicates that different types of entanglement behave in a very different manner. 

Let us now sketch the idea of the protocol for creating Dicke type entanglement; all details are  in Appendix A.2. As in the previous cases, the protocol consists of two steps: a permutation of the diagonal elements followed by a rotation to maximally entangled states (in this case to Dicke states). The permutation first moves the largest eigenvalue, $p^n$, plus the small eigenvalues, $p^k(1-p)^{n-k}$, into the degenerate subspace of energy $kE$, thus purifying the subspace. It also moves other small eigenvalues\footnote{i.e., eigenvalues with population $p^{n-m}(1-p)^m$ with  $m/n \rightarrow 0$ in the asymptotic limit.} into the subspaces of $k-1$ and $k+1$ excitations, as this is favorable for the considered entanglement witness \cite{Dick}. Now, in the degenerate subspace of $k$ excitations, the state with the biggest population $p^n$ is rotated to the Dicke state (\ref{dickeK}). In order for the transformation to be unitary, the rest of the energy eigenvectors of the subspace are rotated to the set of orthonormal states
\begin{equation}
|d^n_{k,l} \rangle = \frac{1}{N_k}\sum_j e^{i\frac{2\pi lj}{N_k}}P_j\{|1\rangle^{\otimes k} |0\rangle^{\otimes n-k}\},\hspace{10mm} N_k=\sqrt{n \choose k} .
\end{equation}
with $i=\{1,...,N_k-1 \}$. This concludes the protocol leading to (\ref{e:TGME}) (see the appendix for detailed computations).

The fact that the creation of Dicke type GME is so much more favorable can be understood intuitively by recalling that Dicke states are in general much more robust to noise compared to GHZ states \cite{Dick,Ali}. Notice also from (\ref{e:TGME}) that it is most favorable to create entanglement in the first excited subspace, where the Dicke state becomes the well-known W state. \blk

\emph{Upper bounds and discussion}

\begin{figure}[t!]
\begin{center}
  \includegraphics[scale=1.5]{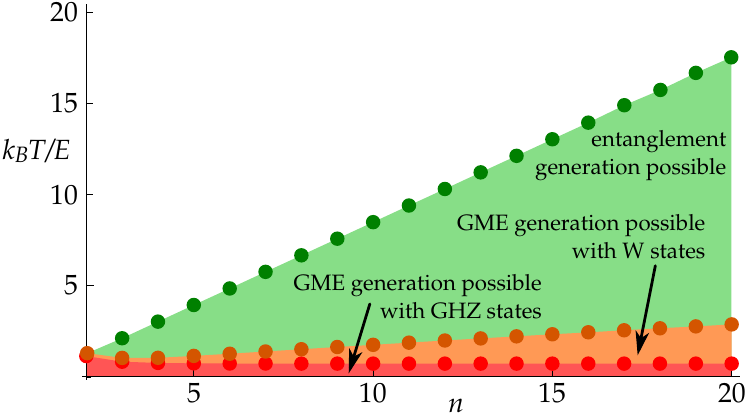}
  \caption{\label{fig2} Regions where entanglement creation (green area) and GME creation (orange area) are possible by our explicit protocols. The upper (green) points represent the best protocol for entanglement creation in a qubit/qudit bipartition, and also an upper bound for creation of GME. The scaling for each region are given in the main text.}
\end{center}
\end{figure}

So far, we have investigated explicit protocols, which allowed us to place lower bounds on the threshold temperature that still allows for the creation of entanglement. To study the limitations imposed by a thermal background it is essential to also find upper bounds on the maximal temperature. For that purpose, a first approach is to use results on the geometry of quantum states. In particular, it is known that the maximally mixed state is always surrounded by a ball of finite size that contains only separable states, and it is possible to place lower bounds on the radius of such a ball \cite{gubaII,gubaIII}. By applying these results we obtain an upper bound that scales exponentially with $n$. Therefore, there is an exponential gap between lower and upper bounds, thus making this approach essentially useless for large $n$. 

The results from \cite{gubaII,gubaIII} are useful for any state, as long as it is sufficiently close to the identity, whereas here we are concerned with a very particular form of states, namely those states with a thermal spectrum. This information can be used to obtain better upper bounds. Indeed, the following theorem was proven in ref.~\cite{Nathaniel}: let $\rho \in \mathcal{H}_2 \otimes \mathcal{H}_d$ have eigenvalues $\lambda_1 \geq \lambda_2 \geq ... \geq \lambda_{2d} $, then  
\begin{equation}
U\rho U^{\dagger} {\rm \hspace{1mm}is \hspace{1mm}separable \hspace{1mm}\forall U}\hspace{1mm} \iff \lambda_1 - \lambda_{2n-1} - 2 \sqrt{\lambda_{2n-2} \lambda_{2n}}\leq 0.
\label{condqudit}
\end{equation}
By taking $d=2^{n-1}$, this criterion applies to any qubit/qudit bipartition of the $n$-qubit thermal system we considered. Furthermore, notice that this condition amounts to calculating the concurrence in a specific $4\times 4$ subspace, which happens to be exactly the purest one we used in the protocol leading to (\ref{T_E}). Hence that protocol is optimal for generating entanglement in any qubit/qudit bipartition. While the possibility to obtain a better $T_\rE$ in a qudit/qudit bipartition remains open, this criterion does yield upper bounds for $T_{\rE}^{\rm(all\hspace{1mm} bip.)}$ and $T_{\rGME}$ \footnote{recall that the presence of entanglement in every bipartition is a necessary condition for GME.}, obtaining 
\begin{eqnarray}
\frac{k_\rB T_{\rE}^{\rm(all\hspace{1mm} bip.)}}{E}\leq \frac{n-1}{\ln 3}, \nonumber\\
\frac{k_\rB T_{\rGME}}{E} \leq \frac{n-1}{\ln 3}
\end{eqnarray}
Therefore we obtain upper bounds on (\ref{Teall}) and (\ref{Tgme}) that also scale linearly with $n$, showing that this scaling between the maximal temperature and the number of qubits is a fundamental property, and that our protocols perform close to optimal for entanglement and GME generation at high temperatures. The results are summarized in fig. \ref{f:concurrence}.

The problem of the attainable entanglement in the unitary orbit of mixed states has been considered in the context of nuclear magnetic resonance (see \cite{chuang} and references therein). The best protocol in ref. \cite{chuang} obtains precisely the scaling (\ref{Teall}), improving on protocols based on algorithmic cooling \cite{Schulman} and effective pure states \cite{Dur}. Our result (\ref{T_E}) provides a tighter bound on the minimal temperature required for entanglement generation, and the upper bound derived from ref. \cite{Nathaniel} gives evidence that it is tight \footnote{Recall that this upper bound only applies for qubit/qudit bipartitions}. We also studied the minimal temperature for GME, finding a surprising positive scaling with the number of qubits.  Our results thus provide bounds on the number of required qubits to generate entanglement and GME at finite temperature, while showing that in the asymptotic limit generation of entanglement and GME is possible at any temperature.

\section{Energy cost}

We can associate to every operation $U$ a work cost $W$, given in (\ref{Wcost}), which corresponds to the external energy input. Regardless of the operation $U$, the invested work is always positive because the initial state is in thermal equilibrium, i.e., $W\geq 0$ $\forall U$. This naturally raises the following question: what is the minimal work cost of correlating thermal state? or, equivalently, what is the maximal amount of attainable correlations when the energy at our disposal, $\Delta E$, is limited?  In this section we address these question, both for total correlations and entanglement, in the unitary orbit of thermal states (i.e., optimizing over all global unitaries $U$). 

\subsection{Correlations}

In analogy with the previous section, let us start by considering the case of a two qudit system, i.e. two $d$-level systems. The goal is now to maximize $I(A_1:A_2)$, as defined in (\ref{MI}), over all global unitaries constrained by $W\leq \Delta E$.

\begin{figure}[h!]
\begin{center}
  \includegraphics[scale=1]{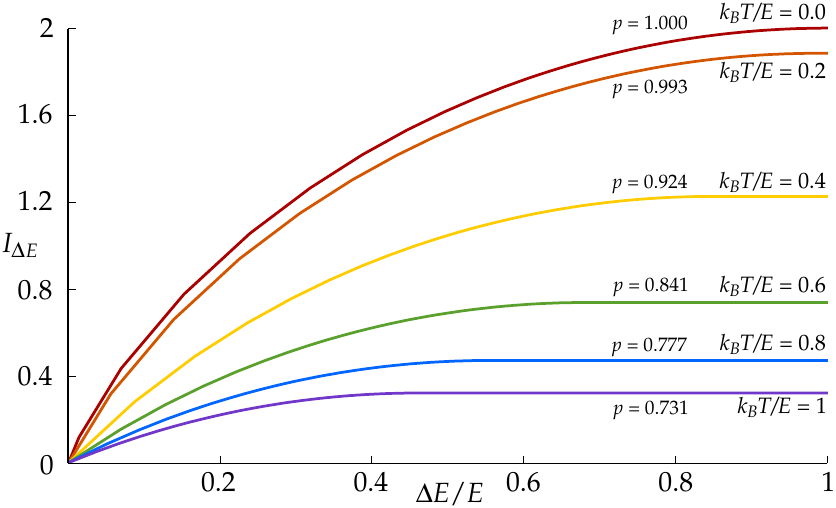}
  \caption{\label{f:MI} Mutual information vs. available energy, for various values of $k_\rB T/E$.}
\end{center}
\end{figure}

Note first that initially $I(A_1 : A_2) = 0$, as the initial state factorizes. Now, to create correlations, we must apply a global unitary that will increase the local entropies $S(A_i)$ of $\rho_f$, since the total entropy $S(A_1A_2) = 2S(\tau_\beta)$ will clearly not change. Recalling that the thermal state maximizes the entropy of a system with fixed average energy (see, for example, \cite{jaynes}), we find that
\begin{equation}
I_{\Delta E} \leq 2\left[S(\tau_{\beta'})-S(\tau_{\beta})\right],
\label{IDeltaE}
\end{equation}  
where $\beta'$ is chosen such that $\Delta E=\Tr[H_{\rm tot} (\tau_{\beta'}^{\otimes 2}-\tau_{\beta}^{\otimes 2})]$. Hence in order to obtain correlations at minimal energy cost, one should look for a protocol such that the local states of $\rho_f$ are thermal states at equal temperature. That is, the optimal unitary $U^{*}$ satisfies 
\begin{equation}
\Tr_{A_1}(U^{*}\rho_i U^{*\dagger})=\Tr_{A_2} (U^{*}\rho_i U^{*\dagger})=\tau_{\beta'}.
\label{Uopt}
\end{equation}
This unitary effectively heats up the system locally, while the global system preserves its entropy. In the appendix (first section) we construct $U^{*}$, thus reaching the bound (\ref{IDeltaE}), for Hamiltonians with equally spaced energy levels and for arbitrary Hamiltonians if the temperature difference is big enough.  In Fig. \ref{f:MI} we illustrate our results for two qubits, for various values of $k_\rB T/E$. Finally, notice that expression  \ref{IDeltaE} recovers the case of maximal correlations, (\ref{IDeltaE}), in the limit $\beta'\to 0$, with a corresponding work cost,
\ba 
W  = 2 \left(\frac{1}{d}\Tr H - \frac{1}{\mathcal{Z}}\Tr He^{-\beta H}\right). 
\ea 

These results are easily extendible to the multipartite case. The generalized mutual information (\ref{Igen}) is maximized (for a given energy cost) by those unitaries that satisfy (\ref{Uopt}) for every local state. 

\subsection{Entanglement}

\subsubsection{Bipartite systems}.

\begin{figure}[h!]
\begin{center}
  \includegraphics[scale=1]{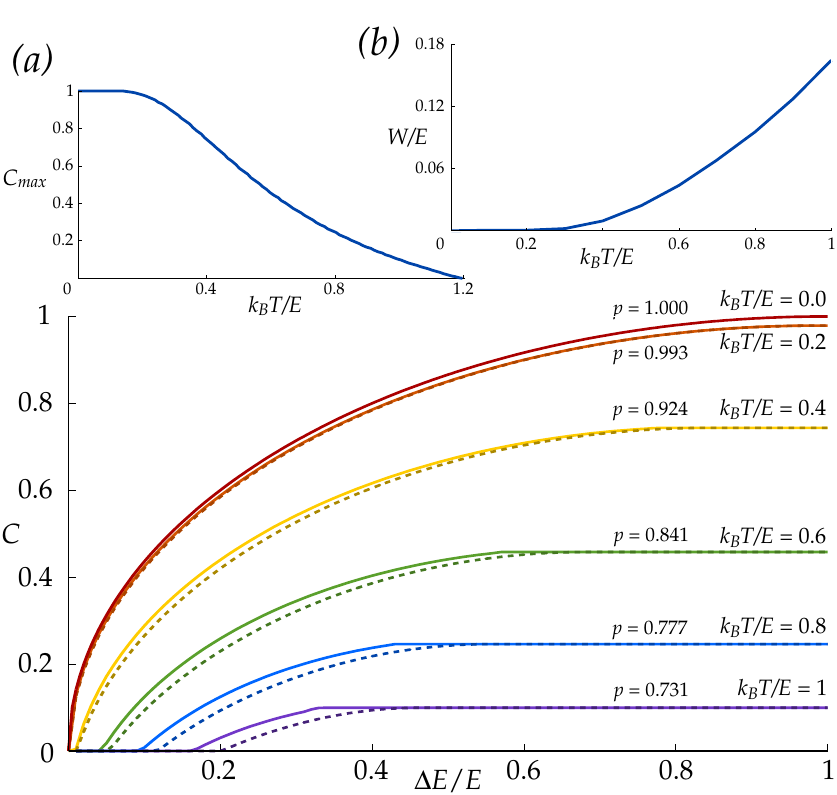}
  \caption{\label{f:concurrence} Main: Concurrence vs. available energy, for various values of $k_\rB T/E$. Solid lines show the optimal protocol, found numerically by optimising over the unitary group. The dashed lines show the performance of the simpler protocol, described in the main text, which is seen to perform well, especially for smaller temperatures. Moreover, if the available energy is not limited, our ansatz is optimal. Inset (a) shows the behaviour of the maximal concurrence $C_\mathrm{max}$ as a function of $k_\rB T/E$, while inset (b) shows the energy needed to leave the separable set, as a function of $k_\rB T/E$. }
\end{center}
\end{figure}

Next we derive the minimal cost of creating entanglement for the simplest case of two qubits. Consider first the case $T=0$, i.e. $\tau = \ket{0}\bra{0}$. If the state is pure, entanglement can be measured by the entropy of entanglement, which is simply given by the local entropy of the state. The problem at hand is thus equivalent to the maximization of the mutual information, so the same reasoning can be used here \footnote{Note that the concurrence and the entropy of entanglement are isomorphic for two qubits}. In particular, the optimal unitary, $U^{*}$ in (\ref{Uopt}), can be generated by a rotation in the $\ket{00}$, $\ket{11}$ subspace. From this we find the relation
\ba 
C=\sqrt{\frac{\Delta E}{E}\left(2-\frac{\Delta E}{E}\right)}.
\ea 
Moving to non-zero temperature, finding the optimal unitary is no longer straightforward. Nevertheless the problem can be attacked from two directions. First, we maximize $C$ numerically, with respect to all possible unitaries, for a given cost $W$. Second, we use an ansatz protocol, inspired by the optimal unitaries to achieve $C_\mathrm{max}$ in (\ref{Cmaxtwoqubits}). These unitaries have the form of first rotating in the subspace of $\ket{10}$ and $\ket{11}$, followed by rotating in the subspace of $\ket{00}$ and $\ket{11}$. Our ansatz is to optimise over such unitaries, now a much simpler optimisation over the two unknown angles (one for each rotation). The results are presented in Fig.~\ref{f:concurrence}, where the solid line shows the result of the full optimisation and the dashed line shows the results of the ansatz. We see that when there is no restriction on the amount of available energy $W$, then our ansatz protocol performs optimally. However, this is not the case when $W$ is limited. Note that the amount of energy required to reach $C_{\mathrm{max}}$ is decreasing as $T$ increases, shown in inset (a), where we also see that for low temperatures ($k_\rB T/E \lesssim 0.1$), we can generate essentially one \red Bell state of two qubits\blk, i.e. $C_{\mathrm{max}} \simeq 1$. Moreover, for any $T>0$, there is a minimal amount of energy required for generating entanglement, shown in inset (b). This is because some energy is always needed to leave the set of separable states. 

\subsubsection{Multipartite systems}.
\textsc{}

Quantification and characterization of multipartite entanglement is still a highly active field of research (see e.g. Ref.~\cite{siewert}). The main challenge is a consistent quantification of multipartite entanglement in operational terms. It seems that this task may not be as easy as in bipartite systems where in the LOCC paradigm entanglement can be quantified by a unique resource. Here we circumvent this issue by studying a measure independent question: what is the energy cost of transforming a thermal state into an entangled one, either GME or entangled in all bipartitions.   

The work cost associated to the unitary (\ref{Ucreation}) is easily computed to be
\begin{equation}
W=\frac{nE(1-e^{-\beta E n})}{2(1+e^{-\beta E})^n}.
\label{Wsep}
\end{equation}
By inserting $T^{\rm all\hspace{1mm}bip.}$ in (\ref{Wsep}), one obtains that the cost to leave the separable set (for this particular protocol) is exactly
\begin{equation}
W^{\rm sep.}=nE\frac{1+\sqrt{2}}{\left(\left(1+\sqrt{2} \right)^{2/n}+1 \right)^n}
\end{equation}
which is exponentially small in $n$. This shows that having more copies not only opens the possibility to generate entanglement at a higher temperature, but also reduces the energy cost of leaving the separable set. An exponential decrease of the work cost with $n$ is also found for the other protocols for GME generation in the multipartite setting (see appendices). The reason behind this behavior is that the considered protocols only act on particular subspaces, whose population becomes negligible in the limit of large $n$. This also implies that the amount of generated entanglement decreases with the number of copies. Interestingly, in the multipartite setting, even a small amount of entanglement might be enough to obtain a substantial quantum advantage. In particular, in the field of quantum computation, for a computational speed up (in pure states) entanglement is required across every bipartition \cite{vidal,josza}, but the actual amount can be polynomially small in the system's size \cite{vandennest} ~\footnote{This translates to density matrices through the convex roof: If every possible decomposition requires at least one element that is entangled across all partitions we can conclude that the classical simulation will be hard and the dynamics of the system non-trivial (while it is not at all clear whether this is necessary it is at least sufficient).}. Therefore, our protocols for multipartite entanglement generation are not only interesting from a fundamental point of view -as they set fundamental bounds on the maximal temperatures- but might find applications in the field of quantum computation.

Finding protocols that generate  a substantial amount of entanglement and GME at high temperatures remains as an important future direction, as this would give a bigger resistance to noise and is important for other applications of GME, such as metrology (\cite{toth}).

\section{Conclusion} 
We have explored the interrelation between two of the most prominent resource theories at the quantum scale: Quantum Thermodynamics and Entanglement Theory. At first we have investigated the impact of imposing entropy constraints arising in thermodynamics on the creation of correlations and entanglement, both in bipartite and multipartite settings. We have worked out fundamental limitations in terms of upper bounds to entangleability, providing necessary conditions for enabling quantum information processing in an unavoidably noisy environment. Furthermore we introduced explicit protocols, and showed that these upper bounds can be (in some cases approximately) reached. In the multipartite setting we studied the advantage of having more systems at one's disposal, providing an explicit route to overcoming some of the fundamental limitations.\\
In a second step we have worked out the energy cost of creating correlations and entanglement, further highlighting the intricate interplay between quantum effects and thermodynamic resources. In the bipartite setting we managed to provide explicit protocols that quantify an upper bound to the work cost of creating a bit or correlation or an e-bit (a fundamental unit of entanglement). We showed that extending the protocols to the multipartite case one can create the strongest form of entanglement at exponentially small energy costs. The introduced protocols serve as ultimate bounds on the possibilities of information processing in scenarios where thermodynamic considerations can not be ignored.\\ 
An interesting open question is the possible implementation of the present protocols in a realistic scenario, which will impose additional restrictions on the class of allowed operations, due to the unconstrained complexity of our introduced protocols. Nevertheless, note that any general unitary can be approximated arbitrarily well in the form of a quantum circuit, involving only single system unitaries and nearest neighbour interactions (gates), with only a small number of distinct interactions required. We leave for future research the exploration of such circuit decompositions, or other physical implementations of our protocols. 

\emph{Acknowledgements.} We would like to acknowledge productive discussions with John Calsamiglia, Daniel Cavalcanti, Paul Erker, Alex Monras and Andreas Winter. M.H. acknowledges funding from the MarieCurie grant N302021 "Quacocos", M.P.L. from the Severo Ochoa program, K.V.H from the Spanish FIS2010-14830 project, N.B. from the Swiss National Science Foundation (grant PP00P2\_138917), P.S. from the Marie Curie COFUND action through the ICFOnest program, A.A. the ERC CoG grant QITBOX.  Part of this work was supported by the COST Action MP1209 ``Thermodynamics in the quantum regime". Parts of it were also carried out at the Quantum Information 2013 conference at Benasque (Spain) and at the Information theoretic approaches to thermodynamics conference at The Institute for Mathematical Sciences, NUS (Singapore).

\begin{appendix}
\section{Appendix}
\subsection{Mutual information in the unitary orbit of thermal product states}
In this appendix we prove that for equal energy spacings ($E_k=k E_0$) any local temperature $T'>T$ lies in the unitary orbit of thermal product states at temperature $T$. Furthermore this proof provides a constructive protocol that in this context reaches any amount of mutual information at minimally possible energy costs. Furthermore for general Hamiltonians this protocols provides a means to reach any sufficiently larger $T'$ (for an exact condition see below) at minimal energy costs. First let us adopt the following convenient notation for the eigenvalue distribution of the marginal $\vec{p}_A:=\textit{diag}(\rho_A)$, which is sufficient for our purposes as the thermal states will always just be diagonal in energy eigenbasis. The general idea of the protocol that follows is that the global unitary should induce a doubly stochastic transformation $M$ on the marginal probability vector, while ensuring that no coherences are created in any subsystem. First we decompose the marginal vectors as
\begin{eqnarray}
\textit{diag}(\rho_A)=\sum_{i=0}^{d-1}\vec{p}_{i}\\
\textit{diag}(\rho_B)=\sum_{i=0}^{d-1}\Pi^i\vec{p}_{i}
\end{eqnarray}
with $p^j_{i}=\langle j|\otimes\langle j+i|\rho|j\rangle\otimes|j+i\rangle=e^{-\beta(E_j+E_{j+1})}/Z^2$ and $\Pi=\sum_k|k\rangle\langle k+1|$. It will be useful to consider again the following generalized Bell states
\begin{eqnarray}
|B_{i,j}\rangle=\sum_{k=0}^{d-1}\omega^{ki}|k\rangle\otimes|k+j\rangle\,,
\end{eqnarray}
with $\omega=e^{2\pi i/d}$. Now it is straightforward to see that rotating in the subspaces spanned by $S_i=span(\{B_{0,i},B_{1,i},(\cdots), B_{d-1,i}\})$ ensures that every diagonal element that can be created by these rotations is being traced over. The unitarity ensures that rotations on the subspaces $S_i$ induce a doubly stochastic transformation of the diagonal part of the density matrix in this subspace. Now all that is left is to observe that the probabilities in the decomposition of the subsystems correspond exactly to the rotations in the subspaces spanned by the maximally entangled states defined before, i.e.
\begin{eqnarray}
\textit{diag}(\textit{Tr}_B(U\rho U^\dagger))=\sum_i M_i\vec{p}_{i}\\
\textit{diag}(\textit{Tr}_A(U\rho U^\dagger))=\sum_i\Pi^i (M_i\vec{p}_{i})
\end{eqnarray}
where each $M_i$ is a doubly stochastic matrix. Now we can use the symmetry of the initial state, i.e. $p_{ij}=p_{ji}$, and define a target doubly stochastic matrix $M$ that should describe the transformation of both marginals. Since the vector $\vec{p}_0$ is equal for both marginals it is evident that $M_0=M$ already takes the first part out of the picture without restricting the generality of stochastic transformations. In general if every doubly stochastic is equal, i.e. $M_i=M$ and commutes with all $\Pi^i$, i.e. is a circulant matrix, it is evident (due to $\sum_{i=0}^{d-1}\vec{p}_{i}=\sum_{i=0}^{d-1}\Pi^i\vec{p}_{i}$) that both subsystems' probability vector will just be transformed by $M$. In other words it is easily achievable to transform the subsystems probability distribution by any doubly stochastic matrix that commutes with all cyclic permutations, i.e. a circulant matrix.\\
The final question is thus whether circulant doubly stochastic transformations of the form $T=\sum_i \alpha_i \Pi^i$ are sufficient to reach any temperature, i.e. $T\vec{p}(\beta)=\vec{p}(\beta')\forall \beta\geq \beta'$? Obviously starting from $\beta=\infty$ one can reach all temperatures via choosing $\alpha_i=p_i(\beta')$ and from any $\beta$ one can reach the infinite temperature distribution via all $\alpha_i=\frac{1}{d}$.
\\
To address this question to its fullest extent we will first construct a general convex sum of cyclic permutations that achieves this general task and then check whether all coefficients are positive. We require that
\begin{eqnarray}
\sum_i\alpha_i\Pi^xp_i(\beta)=p_x(\beta')
\end{eqnarray}
For sake of simplicity we will first define $\alpha_i'=\alpha_i\frac{Z}{Z'}$ such that the condition is simplified to
\begin{eqnarray}
\sum_i\alpha'_ie^{-\beta E_{i+x}}=e^{-\beta' E_{x}}
\end{eqnarray}
A set of $\alpha'_k$ solving this equation system is given by
\begin{eqnarray}
\alpha'_k=\frac{\epsilon_{i_1i_2(...)i_n}e^{-\beta E_{i_1}}e^{-\beta E_{i_2-1}}(\cdots)e^{-\beta' E_{i_n-n-1-k}}}{\epsilon_{i_1i_2(...)i_n}e^{-\beta E_{i_1}}e^{-\beta E_{i_2-1}}(\cdots)e^{-\beta E_{i_n-n-1}}}
\end{eqnarray}
From this explicit form we can easily find negative coefficients and thus prove that circulant matrices are insufficient to reach any arbitrarily higher temperature. On the other hand we immediately see from very simple geometric considerations that for a sufficiently high difference in temperatures $\Delta T=T'-T$ circulant matrices are always sufficient. Since the original probability vector is linearly independent from all its cyclic permutations and all of them are equally far in Euclidean distance from the infinite temperature distribution we can study the convex cone with $\Pi^i\vec{p}_A$ as extremal rays. Since all of the extremal rays share the same distance to the center ray (infinite temperature), we know that a sufficient condition for circulant matrices to achieve the higher temperature Boltzmann distribution is simply given by the minimal distance from the central ray to all faces of the cone. This is always easily calculable for any energy distribution and gives a sufficient condition on $\Delta T$ for this protocol to work.\\
Furthermore we can use the explicit solution to find Hamiltonians for which this protocol always works. One important example is equal energy spacing between the different levels, i.e. $E_k=k E_0$. In this case the explicit solution for the $\alpha_k$ is given as
\begin{eqnarray}
\alpha_k=\frac{1-Z'p_1'}{1-Zp_1}\left( \delta_{0,k} + \frac{Z'p_1'-Zp_1}{1-Z'p_1'}p'_{k+d\delta_{0,k}-1} \right)
\end{eqnarray}
which is positive for all $k$ due to the fact that $p_0>p_0'$, i.e. we have derived a protocol that delivers the maximally possible amount of mutual information at minimum energy costs for all Hamiltonians with equal energy spacing (and thus qubits as a special case).

\subsection{The energy cost and scaling of the W-state protocol}

Given an $n$-qubit thermal state $\Omega=\tau_\beta^{\otimes n}$ we here find the $n\gg1$
asymptotic behaviour  of the maximal temperature $T_{GME}$ that allows to unitarily create genuinely
multipartite entanglement (GME) in the ensemble with the W-state protocol and also calculate the energy cost of the protocol. Here $\tau_\beta=\textit{diag}(p,vp)$,
where $p=1/(1+v)$ and $v=e^{-\beta E}$ is the Boltzmann weight. If the
eigenvectors corresponding to the first excited level of the total Hamiltonian are
$\{ |w^{(1)}_i\rangle \}_{i=1}^n$ and the ones corresponding to the second excited level are
$\{ |w^{(2)}_a\rangle\}_{a=1}^{n(n-1)/2}$, then the measure we use has the form \cite{pauli}
\bea \label{measure}
\mathcal{E}[\Omega]\hspace{-0.75mm}=\hspace{-1mm}\sum_{i\neq j}\hspace{-0.5mm}|\Omega_{ij}|\hspace{-0.5mm}- 2\hspace{-0.2mm}
\sqrt{\Omega_{00}} \sum_{a}\hspace{-1mm}\sqrt{\Omega_{aa}}-\hspace{-0.5mm}(n\hspace{-0.25mm}-2)\hspace{-0.5mm}\sum_i
\hspace{-0.7mm}\Omega_{ii}
\eea
where $\Omega_{ij}=\langle w^{(1)}_i|\Omega|w^{(1)}_j\rangle$ and $\Omega_{ab}=\langle w^{(2)}_a|
\Omega|w^{(2)}_b\rangle$.

In short, the W-state protocol is the maximization of $\mathcal{E}$ over all such unitary
operations that generate non-diagonal elements only in the eigensubspace of the first
excited level (which we denote by $\mathcal{W}_1$). These unitaries can be represented as
$U\Pi$, where $\Pi$ is a permutation operation on the initial state and $U$ is a general
$n\times n$ unitary living in $\mathcal{W}_1$. As this representation suggests, we divide
the optimization procedure in two steps: (i) maximization over $U$s for a given $\Pi$, and
(ii) maximization over $\Pi$s. After $\Pi$ acts, the state becomes $\Omega^\Pi=\Pi\rho\Pi$
and its projection on $\mathcal{W}_1$ we denote by $\omega^\Pi$. Now, the operation $U$
will act only on $\omega^\Pi$ and take it to $\omega'=U\omega^\Pi U^\dagger$ and since $U$
is unitary, the traces of $\omega'$ and $\omega^\Pi$ will be the same. Therefore, we can
rewrite (\ref{measure}) as
\bea \label{aranq1}
\mathcal{E}[U\Pi\Omega\Pi U^\dagger]\hspace{-1.5mm}=\hspace{-1.5mm}\sum_{i\neq j}|\omega'_{ij}| 
                                    \hspace{-1.5mm}-\hspace{-1.5mm}2\sqrt{\Omega^\Pi_{00}} \sum_{\alpha}
\sqrt{\Omega^\Pi_{\alpha\alpha}}-(n-2)\sum_i \omega^\Pi_{ii}.~~~
\eea
This shows that the maximization of $\mathcal{E}$ over $U$ is reduced to the maximization
of $\sum_{i\neq j}|\omega'_{ij}|$ over $U$. To find this maximum, we first observe that
due to the unitarity of $U$, $\Tr\left((\omega')^2\right)=\Tr\left((\omega^\Pi)^2\right)$;
whence,
\bea \label{aranq2}
\sum_{i<j}|\omega'_{ij}|^2=\frac{\Tr\left((\omega^\Pi)^2\right)-\sum_i (\omega'_i)^2}{2}.
\eea
We now relax for the moment the constraint that $\omega'$ and $\omega^\Pi$ are unitarily
connected and only require that $\Tr\left(\omega'\right)=\Tr\left(\omega^\Pi\right)\equiv\alpha$
and $\Tr\left((\omega')^2\right)=\Tr\left((\omega^\Pi)^2\right)\equiv\alpha^2\lambda$.
Here we again divide the optimization in two steps: 1) maximize $\sum_{i<j}|\omega'_{ij}|$
with $\sum_{i<j}|\omega'_{ij}|^2$ fixed and 2) maximize the latter. Now we notice that\\
1) The maximum is reached for $|\omega'_{ij}|=|\omega'_{i'j'}|$ and therefore
$\max\sum_{i<j}|\omega'_{ij}|=\sqrt{\frac{n(n-1)}{2}\sum_{i<j}|\omega'_{ij}|^2}$.\\
2) From (\ref{aranq2}), the maximum for $\sum_{i<j}|\omega'_{ij}|^2$ is reached when
$\sum_i (\omega'_i)^2$ is minimal. Since $\sum_i \omega'_i=\alpha$ is fixed, the minimum
for $\sum_i (\omega'_i)^2$ is reached when all $\omega'_i=\frac{\alpha}{n}$, i.e.,
$\max\sum_{i<j}|\omega'_{ij}|^2=\alpha^2\left( \lambda - 1/n \right)/2$. \\
Finally,
\bea \label{themax}
\max\sum_{i\neq j}|\omega'_{ij}|=\alpha\sqrt{n(n-1)(\lambda-1/n)},
\eea
and on this maximum, $\omega'$ has the following form:
\bea \label{form}
\alpha\hspace{-1mm}\left(\hspace{-1mm} \begin{array}{cccc}
\frac{1}{n} & \hspace{-1mm}e^{\phi_{12}}\hspace{-1mm}\sqrt{\hspace{-1mm}\frac{\lambda-1/n}{n(n-1)}} &
\hspace{-1mm}\cdots\hspace{-1mm} & e^{\phi_{1n}}\hspace{-1mm}\sqrt{\hspace{-1mm}\frac{\lambda-1/n}{n(n-1)}} \\
e^{\phi_{21}}\hspace{-1mm}\sqrt{\hspace{-1mm}\frac{\lambda-1/n}{n(n-1)}} & \hspace{-1mm}\frac{1}{n} &
\hspace{-1mm}\cdots\hspace{-1mm} & e^{\phi_{2n}}\hspace{-1mm}\sqrt{\hspace{-1mm}\frac{\lambda-1/n}{n(n-1)}} \\
& \hspace{-1mm}& \hspace{-1mm}\vdots\hspace{-1mm} & \\
e^{\phi_{N_1}}\hspace{-1mm}\sqrt{\hspace{-1mm}\frac{\lambda-1/n}{n(n-1)}} & \hspace{-1mm}e^{\phi_{N_2}}\hspace{-1mm}
\sqrt{\hspace{-1mm}\frac{\lambda-1/n}{n(n-1)}} & \hspace{-1mm}\cdots\hspace{-1mm} & \frac{1}{n}
\end{array} \hspace{-0.5mm}\right)
\eea
Obviously, being obtained in less restrictive conditions, (\ref{themax}) upper-bounds
the sought $\max_{U}\sum_{i\neq j}|\omega'_{ij}|$. Nevertheless, one can prove, that
for suitably chosen $\{\phi_{ij}\}$ the matrix in (\ref{form}) can always be unitarily
reached from $\omega^\Pi$. The proof is slightly more involved and is conducted by first
proving the statement for $n=3$ by explicitly calculating the corresponding phases
(only one phase is necessary to adjust there) and then proving the statement by
induction for any $n$. \\

Now, having done the maximization over $U$, we turn to finding the $\Pi$ with largest
\bea\label{premax}
\mathcal{E}^\Pi=\alpha\left(\sqrt{n(n-1)(\lambda-1/n)}-n+2\right)
               -2\sqrt{\Omega^\Pi_{00}} \sum_{\alpha}\sqrt{\Omega^\Pi_{\alpha\alpha}},
\eea
where we have plugged (\ref{themax}) in (\ref{aranq1}).
The quantity $\lambda$ is defined above as the sum of the squares of the normalized
elements of $\omega^\Pi$. Therefore, it is never bigger than $1$ which implies that
in the $n\to\infty$ limit, $\mathcal{E}^\Pi$ in (\ref{premax}) will be non-negative only if
$\lambda\to1$. On the other hand, choosing a bigger $\alpha$ and smaller elements in
the eigensubspace of the second excited level (which we denote by $\mathcal{W}_2$) and
on the ground state will also make $\mathcal{E}^\Pi$ bigger. To fulfil all this we
choose $\Pi$ so that it takes the smallest element of $\Omega$, $p^nv^n$, to the ground
state, the biggest one, $p^n$, to $\mathcal{W}_1$. The rest of $(n-1)$ elements in
$\mathcal{W}_1$ are chosen so that they are significantly smaller than $p^n$. We will
take them to be all equal (so that they keep $\alpha$ as big as possible) and to be
$p^nv^{n-k}$ with some $k$ that will be discussed later on. Also, we will choose the
elements in $\mathcal{W}_2$ to be $p^nv^{n-m}$ with some $m$ that is small and
independent of $n$. At this point we do not know which exact choice of $k$ and $m$
will maximize $\mathcal{E}^\Pi$, but fortunately the existing information about
them is enough to deduce the asymptotic behavior we need.

With above described $\Pi$ we have
\bea
\lambda=\frac{1+(n-1)v^{2(n-k)}}{\left(1+(n-1)v^{n-k}\right)^2}.
\eea
So, to have $\lambda\to1$, $nv^{n-k}$ must $\to0$. With this condition and some
algebraic manipulations employing Taylor expansions, we arrive at the following
asymptotic expansion:
\bea \label{asy1}
\frac{\mathcal{E}^\Pi}{p^n}=1-n^2v^{n-k-m/2}C_n+\mathcal{O}\left[n^3v^{2(n-k)}\right],
\eea
where $C_n=(1-1/n)^2\left(v^{m/2}+v^kn/(n-1)\right)$ and is always $\mathcal{O}[1]$ since
$v<1$ and $k$ and $m$ are positive. With this, we rewrite (\ref{asy1}) as
\bea \label{asy2}
\frac{\mathcal{E}^\Pi}{p^n}=1-n^2v^{n-k-m/2}C_n\left(1+\mathcal{O}\left[nv^{n-k}\right]\right).
\eea
Having in mind that $nv^{n-k}\to0$ and explicitly indicating the dependence of $k$
and $T$ (and hence $v$) on $n$ we obtain from (\ref{asy2}) the asymptotic condition
of the positivity of $\mathcal{E}^\Pi$ in the following form:
\bea \label{asy3}
1\geq n^2v_n^{n-k_n-m/2}C_n=e^{\ln C_n+2\ln n-\frac{n-k_n-m/2}{T_{GME}/E}}.
\eea
From formula (\ref{asy3}) it is now obvious that to maximize $T_{GME}$, $k_n$ has to be
as small as possible. So, whatever the $k_n$ and $m$ delivering the maximum are, they
are finite numbers independent of $n$. Therefore,
\bea
T^{\rm max}_{GME}=\frac{nE}{2\ln n}+\mathcal{O}\left[\frac{nE}{(\ln n)^2}\right].
\eea
Finally the energy input required for such a scaling can simply be calculated from the prior permutations $\Pi$ alone, as all subsequent rotations are performed in a degenerate subspace. Adding the cost of all the permutations above gives the rather cumbersome formula for the energy cost of the W-state protocol as
\bea
W=E(1-e^{-\beta E})^{-n}\big((n-1)(e^{-\beta E}-e^{-\beta E(n-1)})+(1-e^{-\beta E})+ne^{-\beta E(n-3)}-\nonumber\\ne^{-\beta E n}+(n^2-n)(e^{-2\beta E}-e^{-(n-2)\beta E})+3(e^{-\beta E n}-e^{-\beta E (n-3)})\big)
\eea 
which while seemingly complicated due to the numerous required permutations still remains exponentially small in $n$ for any $T>0$.

The $W$-state is but an element of a larger set of Dicke states. Correspondingly, our $W$-state protocol can be straightforwardly generalized to Dicke state protocols. First, let us introduce the $m$ excitation Dicke states for $n$ qubits:
\bea
|D_m\rangle=\frac{1}{\sqrt{C_n^m}}\sum_{\{ \alpha \}}|\{\alpha\}\rangle,
\eea
where $\{\alpha\}$s are the subsets of $\{i\}_{i=1}^n$ consisting of $m$ elements, $|\{\alpha\}\rangle=\bigotimes_{i\in\{\alpha\}}|1\rangle_i\bigotimes_{j\in\{ j\not\in\{\alpha\} \}}|0\rangle_j$, and the summation runs over all $C_n^m$ possible $\{\alpha\}$s. Accordingly, the Dicke state protocol is the one when one is allowed to create non-diagonal elements only in $\mathcal{D}_m$ -- the subspace spanned by $|\{\alpha\}\rangle$s. In that case, the GME witness is as follows \cite{pauli}:
\bea
\mathcal{E}_m[\Omega]=\sum_{\{\gamma\}}\left( |\langle\{\alpha\}|\Omega|\{\beta\}\rangle| - \sqrt{\langle \{\alpha\} |\otimes\langle \{\beta\} |\Pi_{\{\alpha\}}\Omega\otimes\Omega\Pi_{\{\alpha\}} | \{\alpha\} \rangle\otimes| \{\beta\} \rangle } \right)-\nonumber\\
n(n-m-1)\sum_{\{\alpha\}}\langle\{\alpha\}|\Omega|\{\alpha\}\rangle,~~
\eea
where the set $\{\gamma\}$ is the collection of all possible $(\{\alpha\},\{\beta\})$ with $\{\beta\}\in\mathcal{D}_m$ and such that the intersection $\{\alpha\}\cap\{\beta\}$ contains $m-1$ elements. As is straightforward to check, $\{\gamma\}$ has $m(n-m)C_n^m$ elements. $\Pi_{\{\alpha\}}$ is a permutation operator which, acted on some $|0...1...0...1\rangle\otimes|1...0...0...1\rangle$, swaps the parts of the vectors corresponding to $\{\alpha\}$ so that it takes first vector to $\mathcal{D}_{m-1}$ and the second one to $\mathcal{D}_{m+1}$; e.g., $\Pi_{\{2,3\}}|01100\rangle\otimes|11000\rangle=|01000\rangle\otimes|11100\rangle$ (see \cite{pauli} for more detailed explanations).

As above, the idea is to maximize $\mathcal{E}_m[U\Pi\Omega\Pi U^\dagger]$ over all unitaries $U$ acting in $\mathcal{D}_m$ and permutations $\Pi$. Again, for a fixed $\Pi$ one has to maximize $\sum_{\{\gamma\}} |\langle\{\alpha\}|\Omega|\{\beta\}\rangle|$, but since $\{\gamma\}$ does not run over all non-diagonal elements the form (\ref{form}) may not necessarily be the optimal one. Nevertheless, since finding maximum of the sum of absolute values of the part of non-diagonal elements of a matrix appears to be a formidable task, we will use the form (\ref{form}) as an ansatz. In what follows we will show that the asymptotic behavior for $T^{\rm max}_{GME}$ following from this ansatz is very close to the optimal one. As in the previous case, the permutation delivering the optimal asymptotics will be the one that puts $p^n$ and $p^nv^{n-k}$ (with $k\geq m$ finite but sufficiently big) in $\mathcal{D}_m$ and fills $\mathcal{D}_{m-1}$ and $\mathcal{D}_{m+1}$ with some $p^nv^{n-l}$ (with sufficiently big and finite $l$). With this, after simple manipulations we arrive at
\bea
\mathcal{E}_m^{\Pi}=p^nm(1+ (C_n^m-1)v^{n-k}-C_n^m(n-m)v^{n-k}-C_n^m(n-m)v^{n-l}+\nonumber \\\mathcal{O}\left(n^{2m}v^{2n}\right) ).
\eea
So the condition for the presence of genuinely multipartite entanglement, $\mathcal{E}_m^{\Pi}\geq0$ reduces to \cite{fooot}
\bea
e^{(m+1)\ln n-\frac{nE}{T}}\leq 1
\eea
whence we obtain
\bea
T\sim \frac{nE}{(m+1)\ln n}
\eea
implying that 
\bea
T^{\rm max}_{GME}\geq \frac{nE}{(m+1)\ln n}
\eea
for large $n$s.

Now, returning to the question of how close to the optimal this scaling is, let us observe that the maximum for $\sum_{\{\gamma\}} |\langle\{\alpha\}|\Omega|\{\beta\}\rangle|$ is given by $\sqrt{(\lambda-1/N)N(N-1)}$ with $N$ in this case being $C_n^m$ (see (\ref{form}) and the reasoning preceding it). This value for $\sum_{\{\gamma\}} |\langle\{\alpha\}|\Omega|\{\beta\}\rangle|$ is not necessarily unitarily achievable from the initial diagonal state but is clearly an upper bound for it. Proceeding as above with this ansatz we obtain 
\bea
T^{\rm max}_{GME}\leq \frac{nE}{m\ln n}
\eea
showing that the initial ansatz (\ref{form}) is quite reasonable and that in any case $T^{\rm max}_{GME}=\mathcal{O}\left(\frac{n}{\ln n}\right)$ for all $m$.

\subsection{Protocols using X-states}

Given a set of $n$ thermal qubits, $\tau_{\beta}^{\otimes n} $, in this section we study the limitations for entanglement creation within unitary transformations of the form:
\begin{equation}
\tau_{\beta}^{\otimes n} \rightarrow{U} \hat{X}
\label{PUU}
\end{equation}
where in the computational basis $\hat{X}$ takes the form
\begin{eqnarray}
\label{Xstates}
\hat{X}=\left(  \begin{array}{cccccccc}
    a_{1} & ¥ & ¥ & ¥ & ¥ & ¥ & ¥ & z_{1} \\ 
    ¥ & a_{2} & ¥  & ¥ & ¥ & ¥ & z_{2} & ¥ \\ 
    ¥ & ¥ & ... & ¥ & ¥ & ... & ¥ & ¥ \\ 
    ¥ & ¥ & ¥ & a_{n} & z_{n} & ¥ & ¥ & ¥ \\ 
    ¥ & ¥ & ¥ & z_{n}^{*} & b_{n} & ¥ & ¥ & ¥ \\ 
    ¥ & ¥ & ... & ¥ & ¥ & ... & ¥ & ¥ \\ 
    ¥ & z_{2}^{*} & ¥ & ¥ & ¥ & ¥ & b_{2} & ¥ \\ 
    z_{1}^{*} & ¥ & ¥ & ¥ & ¥ & ¥ & ¥ & b_{1} \\ 
  \end{array}
\right),
\end{eqnarray}
with $n=2^{N-1}$, $|z_{i}|\le \sqrt{a_{i}b_{i}}$ and $\sum_{i}(a_{i}+b_{i})=1$ to ensure that $\hat{X}$ is positive and normalized (see \cite{marcus} for details). A relevant example of an X-like matrix is,
\begin{equation}
\rho= |GHZ\rangle \langle GHZ | + \frac{\openone}{2^n}
\end{equation}
where $|GHZ \rangle = \frac{1}{\sqrt{2}}(|0 ... 0\rangle + |1 ... 1\rangle)$. As shown in \cite{marcus}, the GME $n$-qubit states of the form (\ref{Xstates}) can be computed by the genuine multipartite  concurrence,
\begin{eqnarray}
\label{CGMEX}
C_{GM}= 2\max\{0,|z_{i}|-w_{i}\}, ~ i=0,1,...,n
\end{eqnarray}
where $w_{i}=\sum_{j\neq i}^{n}\sqrt{a_{j} b_{j}}$. 

We wish to maximize (\ref{CGMEX}) over all $U$ acting on (\ref{PUU}). 
The initial state, $\tau_{\beta}^{\otimes n}$, has no off-diagonals term in the computational basis. It is then advantageous to apply a unitary operation that only generates one off-diagonal term. Indeed, creating off-diagonal terms results into a stochastic transformation of the diagonal terms, thus increasing the $w_{i}$ term in (\ref{CGMEX}) while the $|z_i|$ term depends only on the highest off-diagonal term. On the other hand, given two diagonal elements $a_i$, $b_i$ of $\tau_{\beta}^{\otimes n}$, the biggest off-diagonal term that can be generated by a unitary operation is $|a_i-b_i|/2$, which is obtained by a rotation to the corresponding Bell states. Therefore, the optimal protocol can be thought as a combination of:
\begin{enumerate}
\item Rotate two diagonal elements to Bell states in order to maximize $|z_i|$ in (\ref{CGMEX}).
\item Permute the rest of diagonal elements to minimize $w_{i}$ in (\ref{CGMEX}). This is implemented by setting the elements in decreasing order (in $w_{i}=\sum_{j\neq i}^{n}\sqrt{a_{j} b_{j}}$, having product of biggest with smallest, second biggest with second smallest, etc.).
\end{enumerate}
Step 1 is optimized by acting on the ground state and the most excited state. On the other hand, the thermal state is already ordered to optimize Step 2. The first step leads to $|z_i|=(1-e^{-n\beta \epsilon})/\mathcal{Z}^n$; and since $a_i b_i =e^{-\beta \epsilon n}$, we obtain that  $w_i=(2^n-2) e^{-\beta \epsilon n/2}/\mathcal{Z}^n$. In the limit of large $n$, one easily obtains that $k_BT_{GME}/\epsilon \simeq \frac{1}{2\ln(2)}$.

The previous optimization was done in 2 steps (first maximizing $z_i$ and then minimizing $w_i$). Arguably this is not the optimal approach, as doing a bit worse in step 1 can have a global benefit. While this being true, one can easily convince himself that the differences are of $\mathcal{O}(1/n)$, and thus essentially rotating the ground state with a very excited state and then optimally permuting the rest of diagonal elements, will always lead to  $k_BT_{GME}/\epsilon \simeq \frac{1}{2\ln(2)}$. Any other unitary creating $X$-states from thermal states can not perform better.

\end{appendix}


\begin{thebibliography}{99}
\bibitem{fooot} At this point we use fact that $C_n^m\sim\frac{n^m}{m!}$ as $n\to\infty$.
\end{thebibliography}
\end{document}